\documentclass[%
superscriptaddress,
 amsmath,amssymb,
 aps, physrev,
]{revtex4-2}

\usepackage{graphicx}% Include figure files
\usepackage{dcolumn}% Align table columns on decimal point
\usepackage{bm}% bold math
\usepackage{colortbl}
\usepackage{xcolor}
\usepackage{verbatim}
\begin{document}

%\preprint{APS/123-QED}

\title{\textbf{Real-time quantification of fluid flows around bubbles during directional solidification}}
\author{Bastien Isabella}
\affiliation{%
Université Lyon 1, CNRS\\
Institut Lumière Matière\\
UMR5306, F-69100, Villeurbanne, France
}%

\author{Emma Houllegatte}
\affiliation{%
 Sciences et Ingénierie de La Matière Molle\\
 UMR 7615, ESPCI Paris, PSL Research University, CNRS, Sorbonne Universités\\
 75005 Paris, France
}%

\author{Cécile Monteux}
\affiliation{%
 Sciences et Ingénierie de La Matière Molle\\
 UMR 7615, ESPCI Paris, PSL Research University, CNRS, Sorbonne Universités\\
 75005 Paris, France
}%

\author{Sylvain Deville}
\email{Corresponding author: sylvain.deville@univ-lyon1.fr}
\affiliation{%
Université Lyon 1, CNRS\\
Institut Lumière Matière\\
UMR5306, F-69100, Villeurbanne, France
}%

\date{\today}

\begin{abstract}
Directional solidification of bubbly liquids plays a critical role in shaping the microstructure and properties of many materials, yet the fluid dynamics governing bubble behavior during solidification remain poorly understood. Using cryo-confocal microscopy and particle image velocimetry, we quantify fluid flows around bubbles during solidification of water containing surfactants and tracers. Our results reveal that volumetric expansion dominates fluid motion, with velocities scaling linearly with the solidification rate (1--20$~\mu m/s$), while Marangoni flows---hypothesized to play a key role---are negligible ($< 5~\mu m/s$) under our experimental conditions. Diffusiophoresis and thermophoresis also contribute minimally. These findings challenge existing theoretical models and provide a framework for controlling bubble distribution in solidified materials
\end{abstract}

\maketitle

\section{Introduction} 
\label{sec:introduction}

The dynamics of bubbles during solidification is of fundamental and applied importance across a broad spectrum of disciplines, from environmental sciences where bubbles are trapped in sea and lake ice~\cite{ahn2008co2,gow1977growth} to engineering applications such as metallurgy~\cite{zhang2013nucleation}, where porosity induced by gas entrapment can compromise the mechanical integrity of cast alloys. In materials science, bubbles formed during directional solidification of metallic foams or crystal growth~\cite{li2013bubbles,bouaita2019seed} can either be harnessed to design lightweight structures or become detrimental defects, altering optical and mechanical properties~\cite{ghezal2012observation}. Similarly, in cryopreservation~\cite{bronstein1981rejection, korber1988phenomena}, the formation and distribution of bubbles during freezing critically influence cell viability and tissue structure.

At the heart of these phenomena lies the interaction between bubbles and the advancing solidification front, where complex fluid transport mechanisms come into play. During directional solidification, solutes and dissolved gases segregate at the interface due to their limited solubility in the solid phase~\cite{tiller1953redistribution, tyagi2022solute}, creating steep concentration gradients that can trigger bubble nucleation once critical thresholds are exceeded~\cite{lubetkin1988nucleation, geguzin1981crystallization,isabella2026}. These concentration gradients span over a distance that scales as D/V with D the diffusion coefficient of the solute and V the front velocity and is of the order of a few hundreds of microns for advancing velocities of a few microns per second. These gradients of solute distribution ~\cite{meulenbroek2021competing} and also in temperature \cite{young1959motion, merritt1988migration} may induce gradients in interfacial tension and density which may trigger flows near bubble surfaces~\cite{takagi2011surfactant}, potentially altering bubble trajectories and growth dynamics. Additionally, thermophoresis and diffusiophoresis~\cite{zheng2002thermophoresis, anderson1989colloid} could further influence bubble behavior by driving particles or solutes along thermal or concentration gradients. Yet, despite their theoretical relevance, the relative contributions of these mechanisms—particularly in confined systems where bubbles interact with a moving solidification front—remain poorly quantified experimentally. This discrepancy hinders our ability to refine models of solute segregation, bubble nucleation, and porosity formation---key processes for designing materials with controlled microstructures, from ice-templated ceramics to high-performance alloys.

Direct in situ observations of these mechanisms are lacking, particularly in confined systems where bubbles interact with a moving solidification front. Theoretical models predict significant Marangoni flows during solidification~\cite{meijer2024bubble}, yet experimental validation still awaits. Furthermore, the relative contributions of volumetric expansion---driven by the 9\% density change upon water freezing~\cite{petrenko1999physics}---versus interfacial flows have not been quantitatively assessed.

To resolve these uncertainties, we employed cryo-confocal florescence microscopy coupled with particle image velocimetry (PIV) to quantify fluid flows around bubbles during the directional solidification of water containing surfactants and tracer particles. By systematically varying solidification velocities (1--20$~\mu m/s$) and surfactant concentrations (0.001--1~wt.\%), we isolated the contributions of volumetric expansion, Marangoni flows, and phoretic effects—revealing unexpected dominance of the former.

\section{Materials and Methods} 
\label{sec:materials_methods}

To isolate the individual roles of volumetric expansion, Marangoni flows, and phoretic effects, we designed a model system using water, a surfactant (Tween 80), and fluorescent tracers, solidified under controlled thermal gradients and velocities. Cryo-confocal fluorescence microscopy enabled in situ visualization of bubble dynamics, while particle image velocimetry provided spatially resolved velocity fields---allowing us to disentangle the contributions of each transport mechanism.

\subsection{Materials} 
\label{sub:materials}

The surfactant (Tween 80), the aqueous fluorophore (Sulforhodamine B), and the fluorescent nanoparticle solution (FluoSpheres carboxylate 0.2~$\mu m$, yellow-green 505/515) were obtained from Sigma-Aldrich, FluoTechnik, and Thermo Fisher, respectively.

\subsection{Bubbly liquid and sample preparation} 
\label{sub:sample_preparation}

The bubbly liquids were prepared using microfluidic devices fabricated according to the method previously developed by \emph{Lorenceau et al.}~\cite{lorenceau2006high}. Two varieties of capillaries have been used in the construction of microfluidic devices. These comprise hollow round borosilicate glass capillaries, which possess an inner diameter of 700~$\mu m$, an outer diameter of 870~$\mu m$, and a length of 10~cm, alongside hollow square borosilicate glass capillaries with an inner dimension of 1~mm, a wall thickness of 200~$\mu m$, and a length of 10~cm, sourced from CM Scientific. The constricted morphology of the round capillaries is obtained via the use of a pipette puller (Sutter P-1000 Next Generation Micropipette Puller). Introduction of fluids and air into the corners of the square capillary is achieved with a syringe pump (NE-8000 High Pressure) and a pressure generator (Elveflow), respectively. 

The liquid phase consisted of distilled water combined with $10^{-5}~\text{M}$ Sulforhodamine B to achieve a fluorescent solution, and Tween 80 employed as a surfactant, with concentrations ranging from 0.001 to 1~wt.\% depending on the specific solution. The concentrations of Tween 80 were selected to span both below and above the critical micelle concentration (CMC $10^{-3}~ wt.~\%$ ~at~ $25~^\circ \text{C}$). The minimum concentration (0.001~wt.\%) corresponds to a regime where surfactant molecules are predominantly monomeric, whereas the maximum concentration (1~wt.\%) far exceeds the CMC to investigate the effects of micellar aggregates.

The gaseous phase introduced by the pressure generator is air. By controlling the liquid flow rate between 5 and 20~$\mu L/min$ and maintaining a pressure range from 50 to 200~mbar, solutions of monodispersed bubbles with diameters from 40~$\mu m$ to 800~$\mu m$ can be produced. In this study, bubbles with diameters of less than 100~$\mu m$ were used. Bubbles are thus always smaller than the sample thickness. Introducing a small tilt of the sample (1–2~\textdegree) triggers movement of the bubbles toward the elevated side, revealing that the air bubbles do not stick to the glass slide.

Final solutions were prepared by incorporating 3~vol.\% of particles into the bubble solutions. These fluorescent particles functioned as tracers to monitor fluid flow during the experiment. Identical solutes, bubble sizes, and particle concentrations were employed across all samples to ensure comparable outcomes and to evaluate the impact of solidification parameters and surfactant concentration. The prepared solution was subsequently introduced via capillary action into a rectangular Hele-Shaw cell composed of two glass slides (Menzel, $24~mm \times 60~mm$, thickness 0.13--0.19~mm), separated by two spacers (double-sided adhesive tape), to maintain a constant sample thickness of approximately $100~\mu m$, and sealed on one side with a plastic support and on the other with nail polish to prevent evaporation and leakage.

\subsection{Cryoconfocal microscopy} 
\label{sub:cryoconfocal_microscopy}

\textbf{Principle of the setup and definition of the reference frames}

\begin{figure}
 \centering
 \includegraphics[width=10cm]{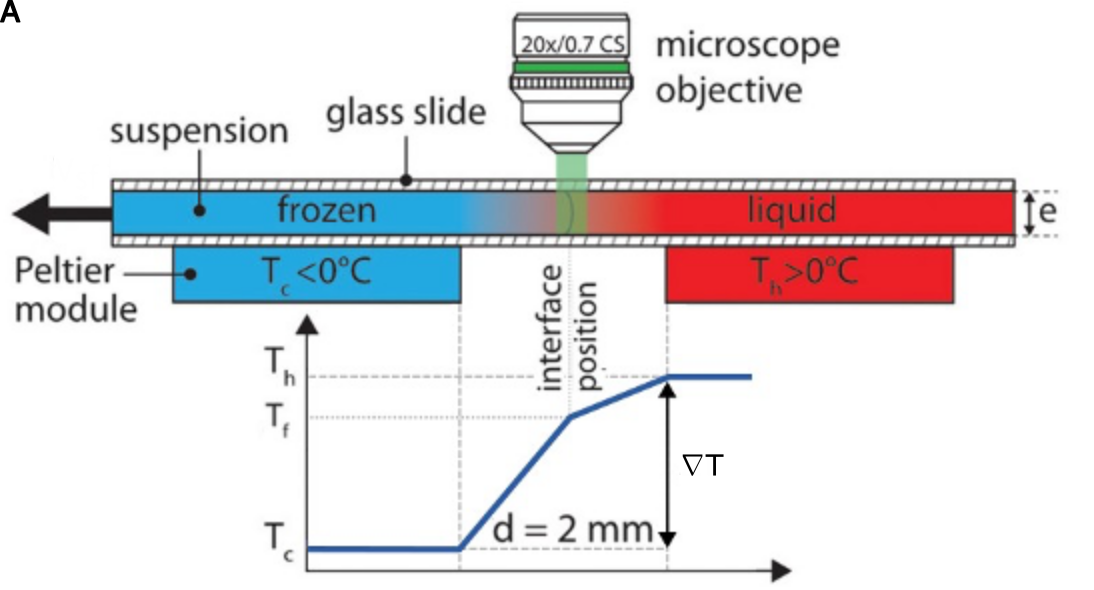}
 \includegraphics[width=6cm]{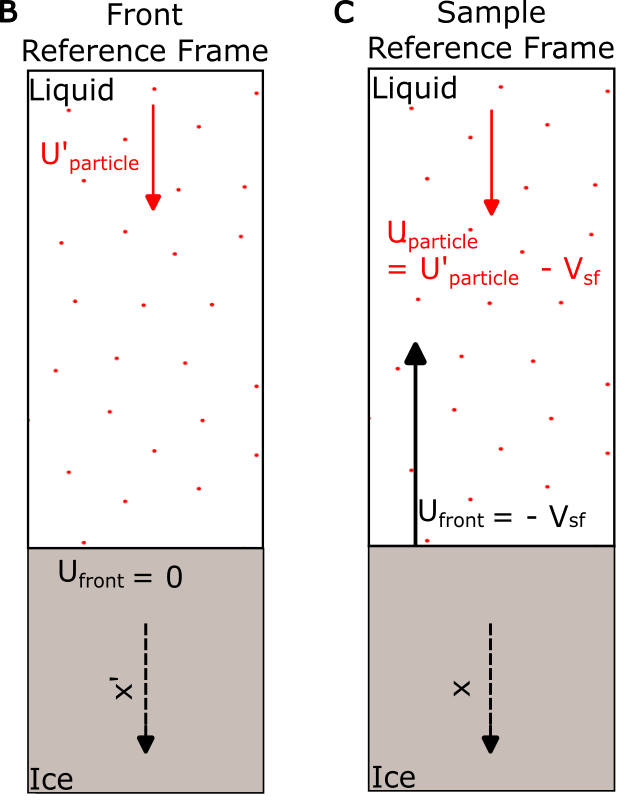}
 \caption{Directional solidification setup and reference frames used for cryo-confocal microscopy experiments. (A) The experimental apparatus is designed to conduct directional solidification experiments using cryo-confocal microscopy. A uniform thermal gradient $\nabla T = T_h - T_c$ is established in the gap $d = 2~mm$ situated between the Peltier modules. The sample is pulled through the temperature gradient at a constant velocity $V_{\text{sf}}$, instigating the growth of ice crystals at a velocity $V_{\text{sf}}$. Consequently, the interface is maintained at a stable position within the observational frame. The temperature $T_f$ is the solidification point of the liquid. (B) A schematic depiction illustrates the reference frames: the sample reference frame and the front reference frame. Microscopy images are captured within the reference frame of the solidification front, wherein the front is stationary. In this frame, bubbles and tracer particles advance toward the front at a velocity $V_{\text{sf}}$, assuming the absence of additional transport mechanisms. The sample reference frame represents the frame in which the solidification front proceeds at a solidification velocity $V_{\text{sf}}$, which is equivalent to the pulling velocity exerted by the motor. Under these conditions, liquid particles remain stationary, provided that no other transport mechanisms are present.}
 \label{fig:figure1}
\end{figure}

We performed unidirectional solidification experiments employing fluorescence confocal cryo-microscopy, using the experimental apparatus previously described~\cite{Dedovets2018-tn} (Fig.~\ref{fig:figure1}A). To summarize, the Hele-Shaw cell was positioned beneath the microscope objective on two Peltier modules, each set at distinct temperatures to generate a thermal gradient across the sample. The sample was advanced at a predetermined velocity $V_{\text{sf}}$ using a stepper motor (Micos Pollux Drive stepper motor with VT-80 translation stage, PI, USA), while the Peltier modules and the microscope themselves remained stationary. By adjusting the Peltier temperatures to be below and above the freezing point, a stationary freezing front is maintained at the solution's freezing point over time. Consequently, images acquired by the microscope are in the reference frame of this stationary front, enabling us to observe the movement of particles and bubbles as they approach the stationary front at a velocity $V_{\text{sf}}$ (Fig.~\ref{fig:figure1}B).

An alternative frame of reference, referred to as the sample frame (Fig.~\ref{fig:figure1}C), can be defined by subtracting the front velocity $V_{\text{sf}}$ from the observed velocities. In this sample frame, the freezing front advances at velocity $V_{\text{sf}}$ through the sample, while the velocities of particles theoretically remain zero at a distance from the front, assuming no interaction occurs between the front and the particles. Consequently, the images are acquired in the front frame, wherein the front appears immobile, and the particles are observed moving toward it. Nonetheless, the velocity data provided in the article are aligned with the sample frame, thus excluding the translational velocity.

\textbf{Thermal gradient and translation stage}

The thermal gradient was established using two Peltier modules controlled by a TEC-1122 Dual Thermo Electric Cooling system from Meerstetter Engineering, Switzerland. Unlike previous studies, enhancements to the experimental setup include enclosing the stage in a Plexiglas box and using compressed dry air to maintain humidity levels between 5\% and 10\%. This modification is intended to prevent condensation on the Peltier modules, thereby reducing any potential friction with the sample and the formation of ice on the samples. The translation velocity $V_{\text{sf}}$ and the temperature gradient can be independently controlled while keeping the solidification front stationary under the microscope objective. Solidification experiments were performed at velocities ranging from 1 to 20~$\mu m/s$ and temperature gradients from 0 to 25$~^\circ \text{C}/mm$.

\textbf{Image acquisition}

Images were acquired using a Leica TCS SP8 confocal laser scanning microscope (Leica Microsystems SAS, Germany), equipped with a long working distance (2.2~mm), non-immersive objective lens (Leica HC PL APO 10x/0.40 CS2). This setup was chosen to provide a sufficiently large field of view while reducing the influence of the microscope's thermal mass on the solidification process. A blue laser with a wavelength of 488~nm was employed in these experimental procedures. Image acquisition was conducted in resonant mode at $512\times 512$ pixels resolution for $582.39\times 582.39~\mu m^2$, leading to a frame rate of 7.5 frames per second for most experiments. An exception was made for the experiment focused on measuring flows in proximity to the bubbles, where images were captured at $512\times 64$ pixel resolution. Two photodetectors were used: one functional within the 504~nm to 516~nm range and the other within the 595~nm to 795~nm range, corresponding to the emission wavelengths of the particles and the Sulforhodamine B dissolved in the liquid, respectively. Air bubbles do not fluoresce and therefore appear black in the images, akin to ice, which also appears black due to the low solubility of solutes within it.

\textbf{Images analysis}

Python and Fiji~\cite{Schindelin2012-lt} were used for image processing and data analysis. Specifically, the flow of liquid within the samples was quantified by tracking the movement of fluorescent particles, using the open-source Particle Image Velocimetry package (OpenPIV~\cite{openPIV}). This methodology enables the detailed mapping of instantaneous vector fields with high spatial resolution, employing fluorescent tracer particles. It offers the ability to visualize and quantify subtle variations in flow around the bubbles and in the vicinity of the solidification front, where gradients in temperature and solute concentration can induce complex flow patterns. The tracers velocity reported in the article are those within reference frame of the sample. They are obtained by subtracting the translation stage velocity, $V_{\text{sf}}$, from the experimentally measured velocity which the one measured in the frame of the solidification front (Fig.~\ref{fig:figure1}).\\

\subsection{Interfacial tension measurements} 
\label{sub:interfacial_tension_measurements}

Surface tension measurements were performed at the interface between bubbles and aqueous solutions containing sulforhodamine B and varying concentrations of Tween 80 through the rising bubble method~\cite{mysels1990maximum}. These measurements were conducted with an automated drop tensiometer and an Interfacial Rheometer (TRACKER\textsuperscript{TM} from TECLIS Scientific). This methodology involves the analysis of bubble morphology, which is extracted and fitted to a Laplace profile. The difference in pressure and interface curvature determined by image analysis provides an estimation of surface tension. Additionally, the effect of liquid temperature was examined using the same methodology; the liquid temperature was regulated using a Minichiller cooler from Peter Huber Kältemaschinenbau SE.

\begin{figure}
 \centering
 \includegraphics[width=8cm]{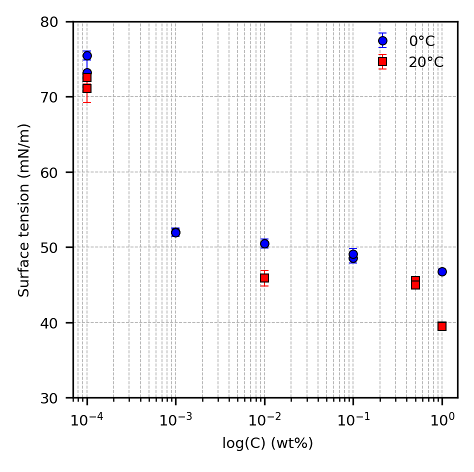}
 \caption{Evolution of the surface tension of bubbles in a solution containing Tween 80 and $10^{-5}~M$ Sulforhodamine B as a function of temperature and Tween 80 concentration C.}
 \label{fig:figure2}
\end{figure}

To assess the potential presence of surface tension gradients induced by temperature gradient, surface tension measurements were carried out across various concentrations C of Tween 80 and under diverse thermal conditions (Fig.~\ref{fig:figure2}). As the temperature decreases, it was observed that there is a consistent increase in surface tension from $45.9 \pm 1.0~mN/m$ at 20~$^\circ$C to $50.5 \pm 0.6~mN/m$ at 0~$^\circ$C, particularly in the case of a surfactant concentration of $C = 0.01~wt.\%$. Moreover, even minimal concentrations of Tween 80 significantly reduce surface tension relative to Sulforhodamine B alone. At 0$~^\circ$C, an additional slight reduction in surface tension is observed as the surfactant concentration increases from 0.001 to 1~wt.\%, with surface tension values decreasing from $52.0 \pm 0.5$ to $46.8 \pm 0.4~mN/m$. As expected, both temperature and surfactant concentration impact the surface tension of the bubble, albeit in different directions. The increase of surfactant concentration near the front should increase the surface tension, while the temperature gradient should have the opposite effect. It is thus difficult to anticipate the overall effect of the combined gradients, and thus the occurrence of Marangoni flows, during directional freezing.

\section{Results and discussion} 
\label{sec:results_and_discussion}

\begin{figure}
 \centering
 \includegraphics[width=14cm]{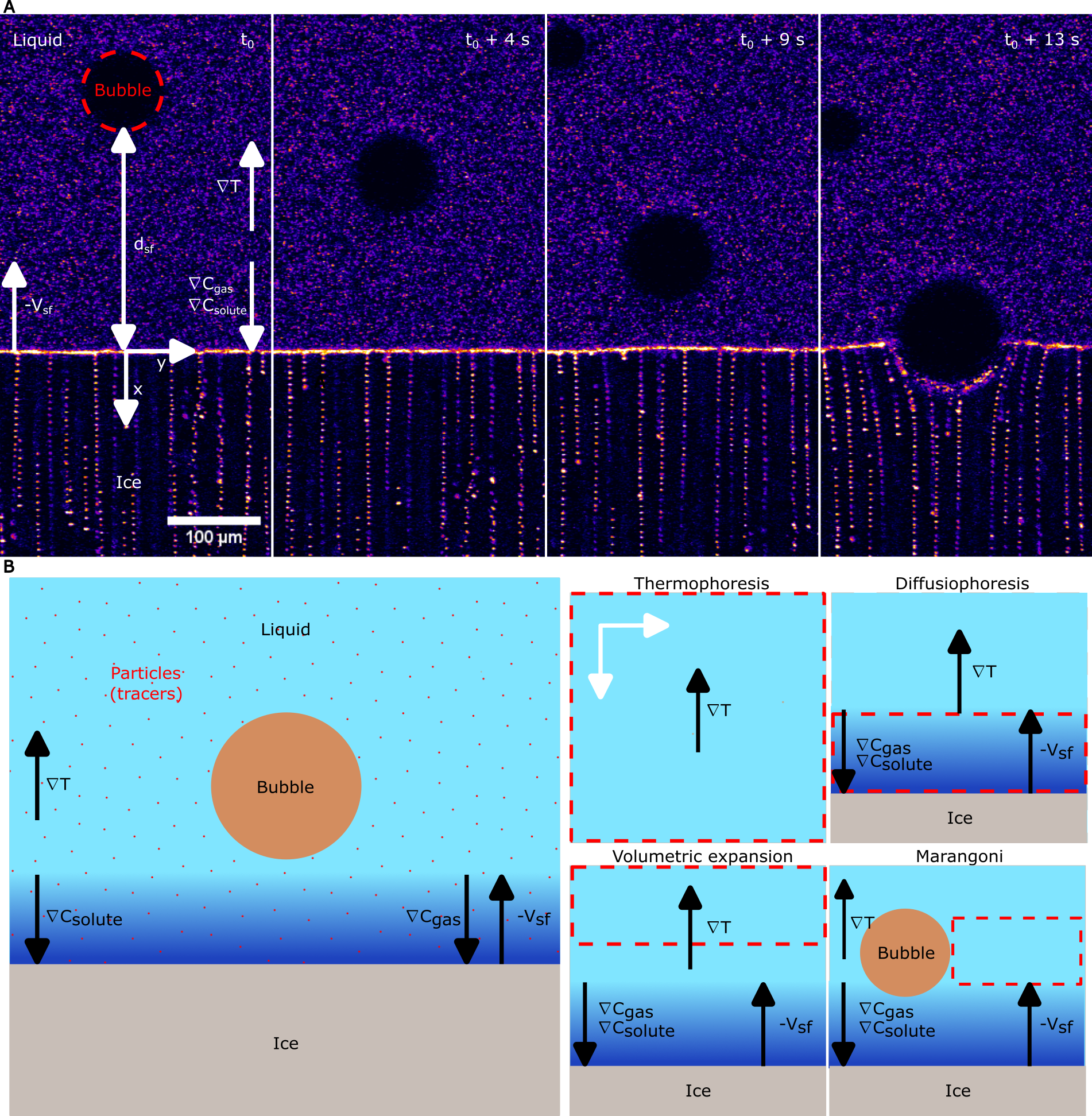} 
 \caption{(A) Example of confocal microscopy images obtained during the freezing of a bubbly liquid. These images are acquired in the reference frame of the freezing front, wherein the front remains stationary, while bubbles and particles move with a defined velocity $V_{\text{sf}}$. $t_0$ corresponds to the time of the first frame of the sequence shown in the panel. (B) A schematic representation of the sample during unidirectional solidification is presented in the sample's frame. Each physical mechanism is methodically investigated via distinct experimental methodologies, as illustrated by the smaller schematics on the right. The regions highlighted in red within these schematics designate areas where flow measurements were conducted.}
 \label{fig:figure3}
\end{figure}

We conducted horizontal solidification experiments by displacing a Hele-Shaw cell containing a mixture of water, bubbles, fluorescent dye, surfactant, and tracer particles using our cryo-confocal stage at a predefined solidification velocity $V_{\text{sf}}$ across a temperature gradient intersecting the freezing point. Fig.~\ref{fig:figure3}A displays a representative sequence of 2D confocal images illustrating the solidification dynamics of our bubbly liquid. In this sequence, the position of the front remains unchanged, while bubbles and particles migrate toward the front, indicating that this image is obtained in the reference frame of the front. To gain insight into the local flow dynamics proximal to the front and bubbles, we incorporated latex tracers into the surfactant solution and used particle image velocimetry. This technique facilitates rapid, comprehensive flow measurements while minimizing the effects of individual particle dynamics, such as Brownian motion. The methodology enables spatially resolved measurements of mean flow velocity, with particular emphasis on its correlation with the distance from the solidification $d_{\text{sf}}$ front and the surrounding bubbles within the liquid phase. The translation velocity, $V_{\text{sf}}$, is subtracted from the measured velocities to present results within the reference frame of the sample, wherein the front velocity advances within the sample at a velocity $V_{\text{sf}}$, and the particle velocity should equal zero in the absence of interactions between the front and the particles.

As discussed in the introduction, the propagation of a freezing front within a suspension of bubbles and surfactants is anticipated to activate multiple mechanisms that impact bubble displacement and fluid dynamics within the sample. It is imperative to disentangle these effects, as illustrated in Fig.~\ref{fig:figure3}B. We list these mechanisms below and will subsequently discuss the conditions under which these effects may be independently assessed.

As a bubble approaches the solidification front, it experiences a gradual reduction in temperature and an increase in surfactant and soluble gas concentration due to their segregation near the solidification front. Such variations in temperature and solute concentration can create interfacial tension gradients along the bubble's surface, potentially leading to convection flows due to density gradients as well as thermal and solutal Marangoni flows in the surrounding liquid. Meijer \emph{et al.}~\cite{meijer2024bubble} have theoretically shown that the density convection due to gas segregation close to the front is negligible but postulated the existence of such thermal and solutal Marangoni flows during the solidification of an aqueous solution containing gas bubbles. However, this hypothesis has yet to receive experimental validation.

Various transport mechanisms may play a significant role in influencing the dynamics of tracer and liquid flow movements. It should be underscored that tracer particles inherently undergo Brownian motion within a liquid medium, independent of any imposed thermal gradient or the introduction of surfactants, gases, or bubbles. The presence of both thermal and concentration gradients may induce thermophoretic and diffusiophoretic movements of the tracer. Additionally, the solidification process may induce flow due to the volumetric expansion that is associated with the solidification of water. For pure water, this expansion is approximately 9\%, demonstrating that the ice formed occupies a greater volume than the liquid state~\cite{petrenko1999physics}. Should this expansion be predominantly oriented along the solidification axis, it may lead to the displacement of liquid in that direction.

In order to differentiate between the various effects under investigation, experiments were conducted under a range of conditions, as documented in the appendix. To suppress the influence of solute segregation close to the ice front, a first set of experiments were performed with a temperature gradient applied across the samples for temperatures exceeding the freezing point in the absence of bubbles (section F). In such conditions only the thermophoretic motion of tracers or micelles is expected to occur.

In order to probe the influence of volumetric expansion during the solidification of water, we conducted freezing experiments without bubbles nor surfactant, specifically applying a temperature gradient that spans the freezing point. To avoid any effect due to gas accumulation close to the freezing front, we conducted velocity measurements at regions in the image that were located more than 150~$\mu m$ away from the solidification front where the solute concentration is expected to be constant. These measurements far from the front were also made in the presence of surfactants and/or bubbles to investigate the influence of thermal gradients and possible thermal Marangoni convection or thermophoretic effect.

Subsequently, to assess the influence of gas and surfactant accumulation, we conducted experiments under identical conditions where the flow measurements were restricted to the areas within 100~$\mu m$ closest to the front. This region correlates with the segregation distance of solutes at the interface, particularly gases and surfactants.

In section~\ref{sub:fluid_flow_close_to_bubbles_to_probe_marangoni_flows}, the goal was to evaluate more specifically the presence of Marangoni close to the surface of the bubbles. Therefore experiments were conducted in the presence of bubbles and the flow measurements were taken exclusively close to the equator of the bubbles as a function of the distance to the bubble surface and to the solidification front. 

The table in the appendix provides a comprehensive summary of all the conditions tested for the study. For instance, the comparative analysis between cases 1 and 4 facilitates the estimation of tracer thermophoresis, based on variations observed in the flow. Likewise, the comparison between cases 4 and 7 elucidates the role of volumetric expansion during the process of solidification.

\subsection{Flows in the presence of a temperature gradient without freezing} 
\label{sub:flows_in_the_presence_of_a_temperature_gradient_without_freezing}

\begin{figure*}
 \centering
 \includegraphics[width=16cm]{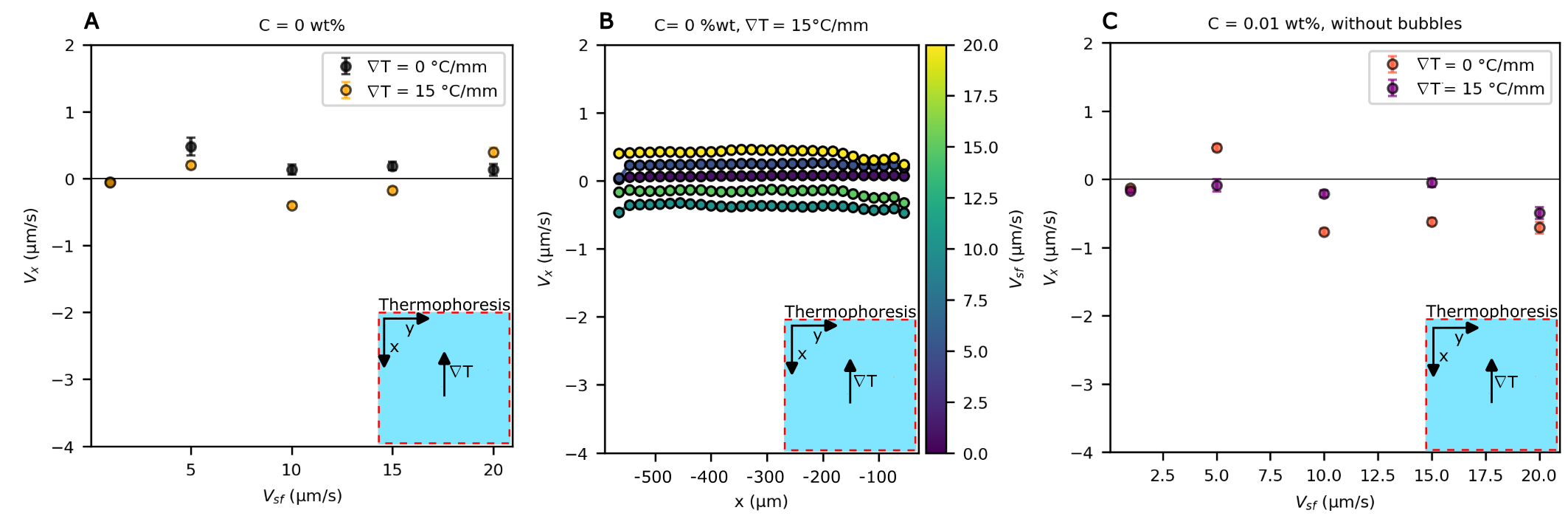}
 \caption{Fluid flow velocity $V_{x}$ in the direction of solidification x as a function of translation velocity $V_{\text{sf}}$ and position in the sample x. (A) Impact of temperature gradient on mean fluid flow velocity in all the sample for a liquid without surfactant. Comparison of cases 1 and 4 from Table 1. (B) as a function of position in the sample (C) in the presence of 0.01~wt.\% of Tween 80. The regions highlighted in red within these schematics designate areas where flow measurements were conducted.}
 \label{fig:figure4}
\end{figure*}

In Fig.~\ref{fig:figure4}, PIV is used to quantify the velocity under the influence of a temperature gradient $\nabla T = 15~^\circ \text{C}/mm$ with Peltier temperatures set at 3$~^\circ$C and 33$~^\circ$C. Fig.~\ref{fig:figure4}A presents the mean velocity calculated across the entire sample, which is subsequently compared to the velocity observed under isothermal conditions in the absence of surfactants. In Fig.~\ref{fig:figure1}B, this velocity $V_x$ is also analyzed as a function of spatial position within the sample, thereby correlating to varying temperatures. In Fig.~\ref{fig:figure1}C, the mean velocity is evaluated in the presence of 0.01~wt.\% Tween 80, which exceeds the critical micellar concentration (CMC).

In the absence of surfactant, the only motion that can be expected is the thermophoresis of the latex particles. Conversely, when surfactant is present, the thermophoretic motion of the surfactant micelles may also induce a motion of the particles. It is observed that, under all three experimental conditions, the mean flow velocities of the sample remain negligible throughout the entire sample.

Thermophoresis is primarily governed by two critical parameters: the temperature gradient $\nabla T$ and the Soret coefficient $ S\textsubscript{T} = D\textsubscript{T}/D$, with D representing the diffusion coefficient, and $D\textsubscript{T}$ the thermophoretic mobility, which is influenced by the temperature of the liquid \cite{piazza2008thermophoresis}. This parameter $D\textsubscript{T}$, which depends on liquid temperature, as well as the type and size of particles, is essential for accurately predicting the thermophoretic velocity $V_T$ of particles under specified experimental conditions \cite{degroot2013}: 

\begin{equation}
V_{T} = \ - \ D_{T}\mathrm{\nabla}T = \ - DS_{T}\mathrm{\nabla}T  
\end{equation}

Consequently, for particles and liquid temperatures analogous to those examined, the Soret coefficient has been measured within the range of -1.5 to 1~${^\circ \text{C}^{-1}}$. This yields anticipated flow velocities $V_x$ less than 0.1~$\mu m/s$~\cite{duhr2006molecules, braibanti2008does}. Hence, the expected thermophoretic velocities are minimal under our experimental conditions, corroborating our observations that suggest the temperature gradient alone does not exert a significant influence on the liquid flows.\

In the presence of surfactants, we did not measure any significant variation of the velocities hence micellar thermophoresis can be excluded as a significant factor in liquid flow dynamics. Consequently, we did not observe any indication of thermophoresis affecting the transport of particles and micelles. 
Therefore, if thermophoresis impacts liquid flows, its influence is exceedingly limited (fluid flow velocity $V_x$ never exceeding 1~$\mu m$/s), with the resulting flows mostly falling below the precision of the motor. From these observations, we infer that under freezing conditions with a temperature gradient of similar magnitude, thermophoresis is unlikely to meaningfully influence the liquid flows observed during the solidification of bubbly liquids.

\subsection{Flows in the presence of freezing} 
\label{sub:flows_in_the_presence_of_freezing}

\textbf{Average velocity measurements far from the front (d\textsubscript{sf}\textgreater 150~$\mu m$) to probe the influence of volumetric extension}

\begin{figure}
 \centering
 \includegraphics[width=8cm]{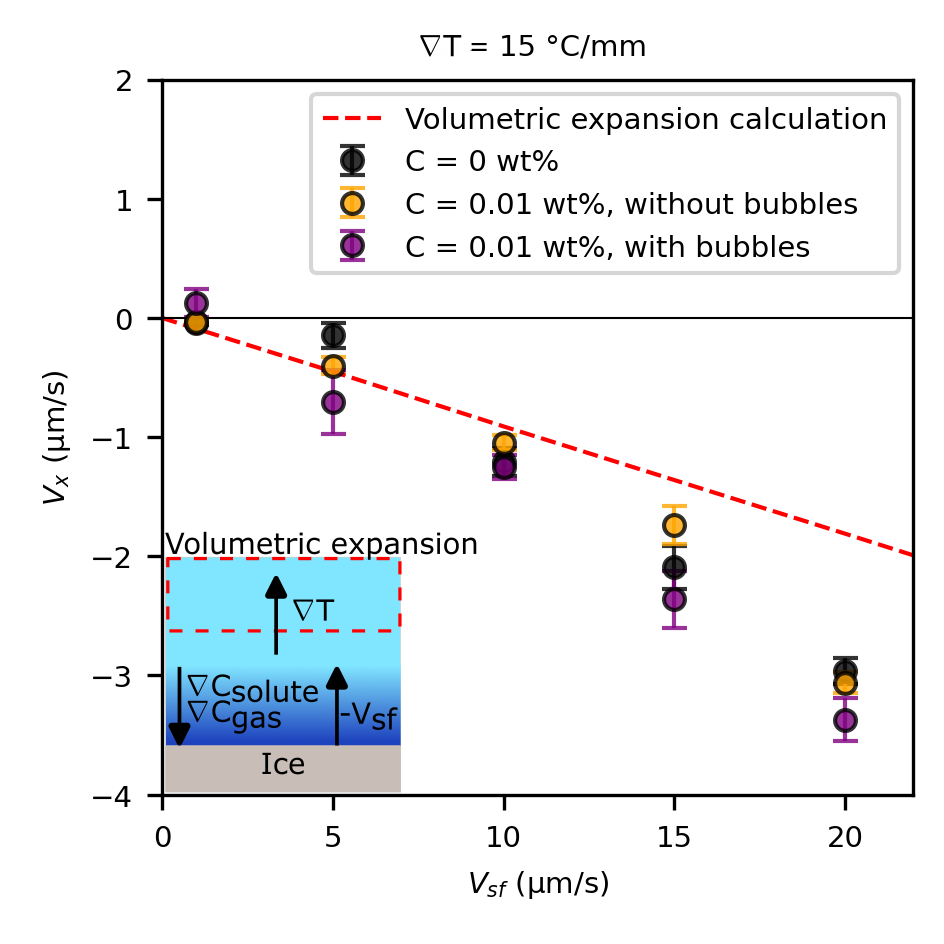}
 \caption{Mean fluid flow velocity $V_x$ measured in the zone located at distances greater than $150~\mu m$ from the solidification front. Negative velocities denote flow oriented in the same direction as the solidification processes. The applied motor velocity is subtracted from the measured to eliminate the effect of sample translation in the observation frame. A red dashed line illustrates the expected effect of volumetric expansion on liquid flows. Comparison between case 7, 8 and 9 from Table~\ref{tab:conditions}. The regions highlighted in red within these schematics designate areas where flow measurements were conducted.}
 \label{fig:figure5}
\end{figure}

In this section, we examine the conditions under which freezing occurs when a temperature gradient of 15$~^\circ \text{C/mm}$ is applied, with temperatures set at 15~$^{\circ}$C and -15~$^{\circ}$C on the Peltier modules (Fig.~\ref{fig:figure5}). We conduct a comparative analysis of experiments performed in the absence of bubbles and surfactants against those conducted with both bubbles and surfactants present. Building upon our previous findings that demonstrated the negligible effect of thermophoretic motion of the particles, we anticipate investigating the impact of volumetric expansion without the influence of bubbles and surfactants.

Furthermore, the measurements are taken far from the front, specifically beyond $150~\mu m$, to mitigate any influence arising from concentration gradients of solutes close to the front. As indicated in reference~\cite{tyagi2022solute}, the distance over which solutes are segregated due to rejection by the front is proportional to $\lambda \approx D/V_{sf}$, where $D= 10~\mu m.s^{-2}$ represents the diffusion coefficient of the micelles. By considering the minimum solidification velocity investigated, $V_{\text{sf}} = 1~\mu m$/s, it is determined that $\lambda = 30~\mu m$ represents the maximum distance within which segregation of the surfactant micelle is expected in surfactant solutions, regarding the solidification velocities under investigation. Conducting measurements at $d_{sf} = 150~\mu m \gg \lambda$ from the front is anticipated to suppress effects attributed to gas and surfactant concentration gradients, such as diffusiophoretic motion or solutal Marangoni flows, particularly when bubbles are present. 

The fluid flow velocity is determined as a function of the solidification velocity $V_{\text{sf}}$. As explained earlier we define a positive velocity $V_x$ when the flow directions are aligned towards the solidification front, thereby resulting in the particles approaching the front. Conversely, a negative velocity means that the distance between the front and the particles increases.

In Fig.~\ref{fig:figure5}, it is evident that an increase in the velocity of the solidification front corresponds to velocities that reach increasingly negative values. This indicates that, during the phase transition from liquid water to ice, the residual liquid is pushed by the solidification front as a consequence of the volume expansion accompanying the phase change. 

The impact of volumetric expansion on liquid flow was estimated by postulating that the entire volume increase occurs ahead of the solidification front and aligns with the direction of solidification, resulting in the liquid being displaced in the same direction. It is assumed that the thickness of the sample remains unchanged upon solidification, and that the sample's lateral dimensions remain constant. In this estimation, it is assumed that the total volumetric expansion associated with solidification is directed entirely along the front progression path. Consequently, this effect can be computed based on the velocity of solidification $V_{\text{sf}}$, as well as the densities of water $d_{\text{water}}$ and ice $d_{\text{ice}}$, using equation 2.
 
\begin{equation}
V_{\text{volumetric}} = V_{\text{sf}} \left( \ 1 - \frac{d_{\text{water}}}{d_{\text{ice}}} \right) 
\end{equation}

The analysis reveals a linear relationship between the solidification rate and the measured liquid flow. This calculation is consistent with our experimental results concerning the solidification velocity as illustrated in Fig.~\ref{fig:figure5} and reported in $V_{\text{sf}} \leq  10~\mu m/s$. Nevertheless, our measurements for $V_{\text{sf}} = 15~\mu m/s$ and $V_{\text{sf}} = 20~\mu m/s$ indicate an acceleration in motion that surpasses anticipated values.  

In the cases where bubbles are present, thermal Marangoni flows may be anticipated as a result of surface tension variations along the bubble surface due to temperature gradient. However the observation of similar velocities in scenarios both with and without surfactants and bubbles allows us to rule out a significant influence of thermal Marangoni flows.

\textbf{Velocity measurements close to the front (d\textsubscript{sf} \textless 150$\mu m$) to probe the influence of solute concentration gradients}\\

We performed velocity measurements closer to the front at a distance $d_{sf} < 150~\mu m$ to the solidification front to assess the potential influence of surfactant and gas concentration gradients, particularly in the proximity of the solidification front. This phenomenon, which may occur in the presence of concentration gradients of smaller solutes, can become operative near the solidification front, where these solutes such gases, dyes, or surfactants, accumulate due to their limited solubility in ice. Given that volumetric expansion is also anticipated under these conditions, we subtracted the velocities measured at a distance from the front from those obtained at $d_{sf}< \lambda \approx 0 ~\mu m$ from the front. The flow velocity $V_x$ thus writes:  

\begin{equation}
    V_x = U'_{\text{particle}} - V_{\text{sf}} -|V_{\text{vol}}|
    \label{eq:velocity}
\end{equation}

No significant velocities were detected near the front. Notwithstanding, the substantial accumulation of tracers at the liquid-solid boundary complicates the precise analysis of flow velocities via this methodology. To address this limitation, employing individual tracer tracking would be advantageous as it facilitates a detailed examination of the impact of solute concentration gradients on particle movement close to the solidification front. This approach will be the subject of an upcoming paper.

\subsection{Fluid flow close to bubbles to probe Marangoni flows} 
\label{sub:fluid_flow_close_to_bubbles_to_probe_marangoni_flows}

Meijer \emph{et al.}~\cite{meijer2024bubble} have theoretically predicted the occurrence of thermal and solutal Marangoni flows during the freezing of bubbles, attributed to temperature gradients and gradients in soluble gas concentration near the interface, anticipated to locally alter the surface tension along the bubble surfaces. Nonetheless, no experimental evidence supporting the existence of such flows has been presented so far. Consequently we investigate in this Section the local flows adjacent to bubbles to test these theoretical predictions.

\begin{figure*}
 \centering
 \includegraphics[width=14cm]{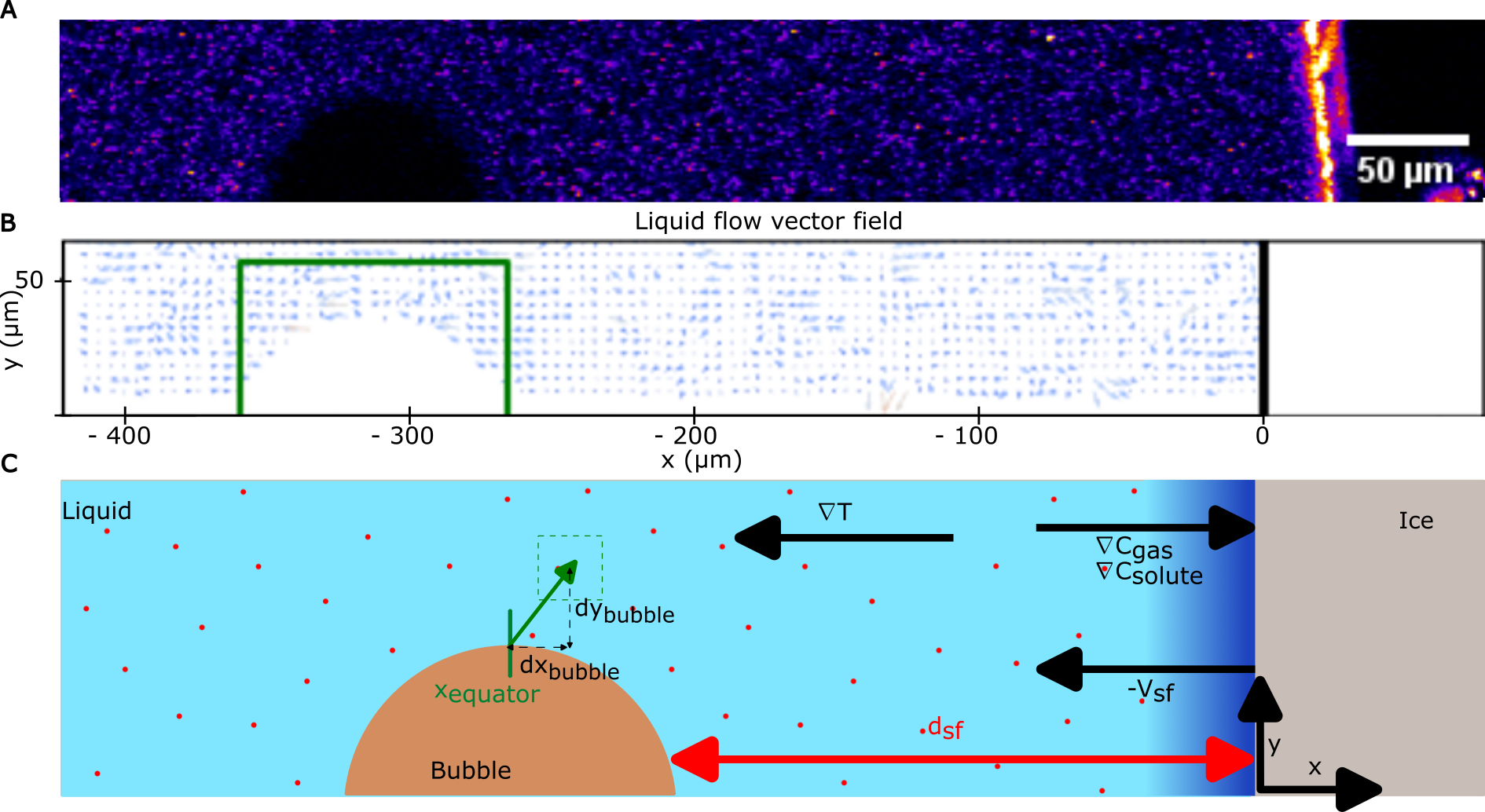}
 \caption{(A) A typical two-dimensional cryo-confocal image from a solidification experiment  used to analyze flow fields at the equator of bubbles in the presence of 0.01~wt.\% Tween 80. (B) Vector fields derived from cryo-confocal microscopy images. The bubble is identified and delineated to eliminate residual vectors that are located inside it. A virtual green box is delineated around the bubble of interest to establish a region of interest for vector analysis. The velocity vectors displayed here represent the measured particle velocities $V_m$, from which the contributions from motor-induced velocity and volumetric expansion $V_{vol}$ have been subtracted $V_x = U'_{\text{particle}} - V_{\text{sf}} -|V_{\text{vol}}|$. (C) A representative diagram of the sample flow around bubbles in the sample reference frame. These images facilitate the identification of the bubble's equator, allowing precise tracking of the flow both at the equator and as a function of the distance from the bubble and from the solidification front.}
 \label{fig:figure6}
\end{figure*}

In order to quantify these flows, reduced observation windows were used to focus the measurements in the vicinity of the bubbles, as depicted in Fig.~\ref{fig:figure6}A. PIV was conducted to obtain the velocity field of the flow. Furthermore, the position of the bubble equator was identified for each frame, enabling the examination of flow at the equator as a function of the distance from the bubble and the distance from the solidification front at each temporal instance (Fig.~\ref{fig:figure6}B). It should be noted that the velocities $V_x$ have been adjusted to account for the imposed motor motion to operate within the sample reference frame. Additionally, the effect of volumetric expansion has been subtracted, as detailed in Fig.~\ref{fig:figure6}.

\begin{figure*}
 \centering
 \includegraphics[width=14cm]{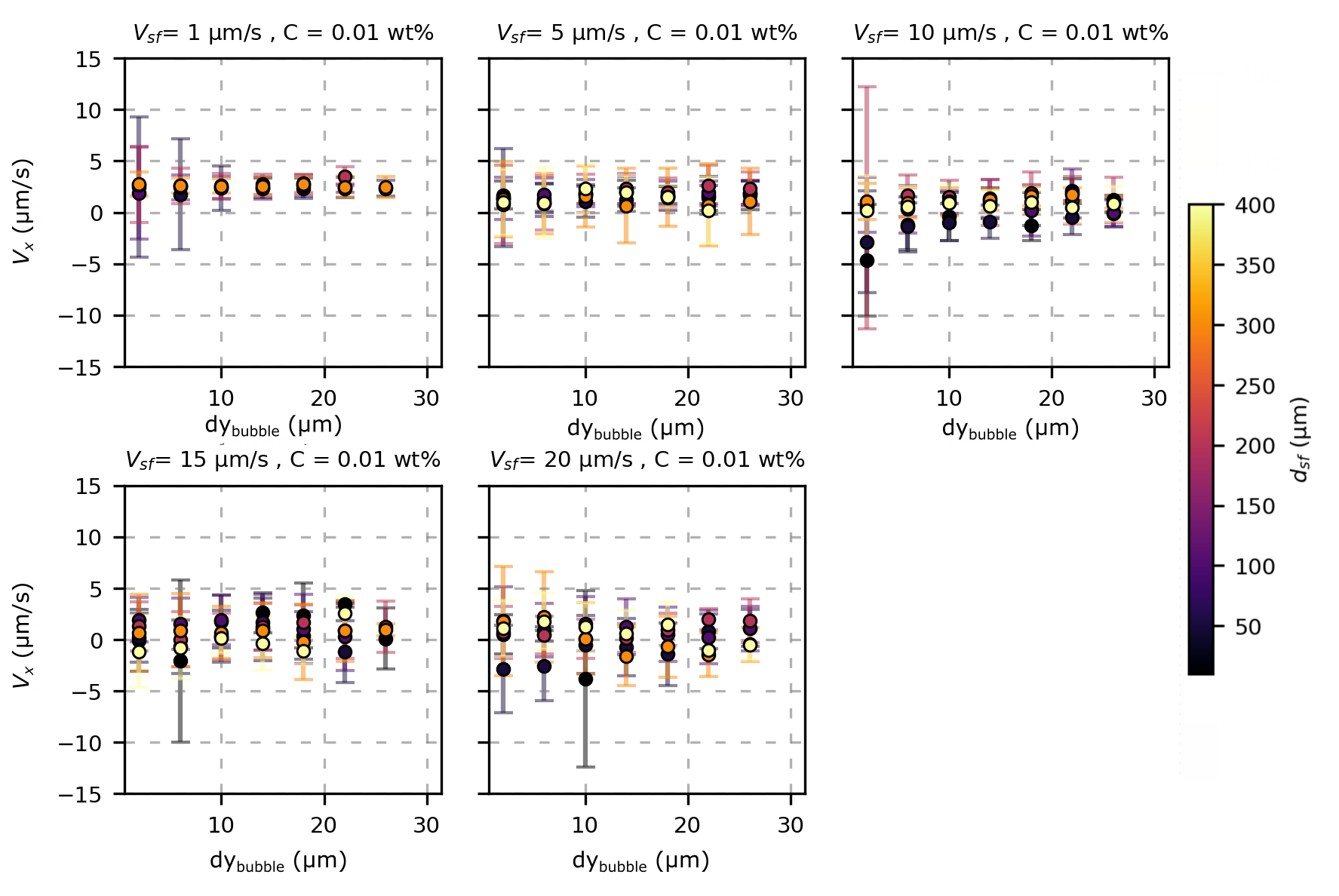}
 \caption{Mean fluid flow velocity $V_x$ near bubbles during liquid solidification, in the direction of the solidification progression, after subtraction of the flow induced by volumetric expansion and the solidification front velocity. The flow is measured at the equator of the bubble within a $\pm 3~\mu m$ range around the $x_{\text{equator}}$ position and is presented as a function of both the distance to the solidification front $d_{\text{sf}}$ and the distance from the bubble $y_{\text{bubble}}$. All experiments were performed using aqueous solutions containing $C = 0.01~wt.\%$ Tween 80 and with solidification velocities $V_{\text{sf}}$ of 1, 5, 10, 15, and $20~\mu m/s$.}
 \label{fig:figure7}
\end{figure*}

In Fig.~\ref{fig:figure7}, we illustrate the flow velocity $V_x$ along the direction of solidification as a function of the distance $d_{\text{sf}}$ between the solidification front and the bubble's leading edge, under the condition of a surfactant concentration of 0.01~wt.\%. This investigation considers solidification velocities $V_{\text{sf}}$ within the range of 1 to 20~$\mu m/s$. The reference for the zero distance point to the solidification front is established when the bubble's leading edge makes contact with the solidification front. Flow values, corresponding to particular distances from the solidification front, were calculated as averages over a $\pm 10~\mu m$ interval centered on the indicated position.

At a solidification velocity of $1~\mu m/s$, the velocity values exhibit minimal variation with respect to the distance from the bubble and from the front; however, the variability of the measured vectors, as indicated by the standard deviation, increases as the distance from the bubble decreases. With higher solidification velocities, such as 5 or 10$~\mu m/s$, the average flow velocities take slightly negative values, ranging from -1 to -5~$\mu m/s$, yet the dispersion of data points increases, and no discernible pattern emerges concerning varying distances from the front and the bubbles. At elevated solidification rates (e.g., 15 and 20~$\mu m/s$), the dispersion of measured flow velocities becomes more pronounced. Furthermore, the mean flow values remain relatively low, within the range of -5 to 5~$\mu m/s$, denoting flows both along and opposed to the solidification direction.

\begin{figure*}
 \centering
 \includegraphics[width=14cm]{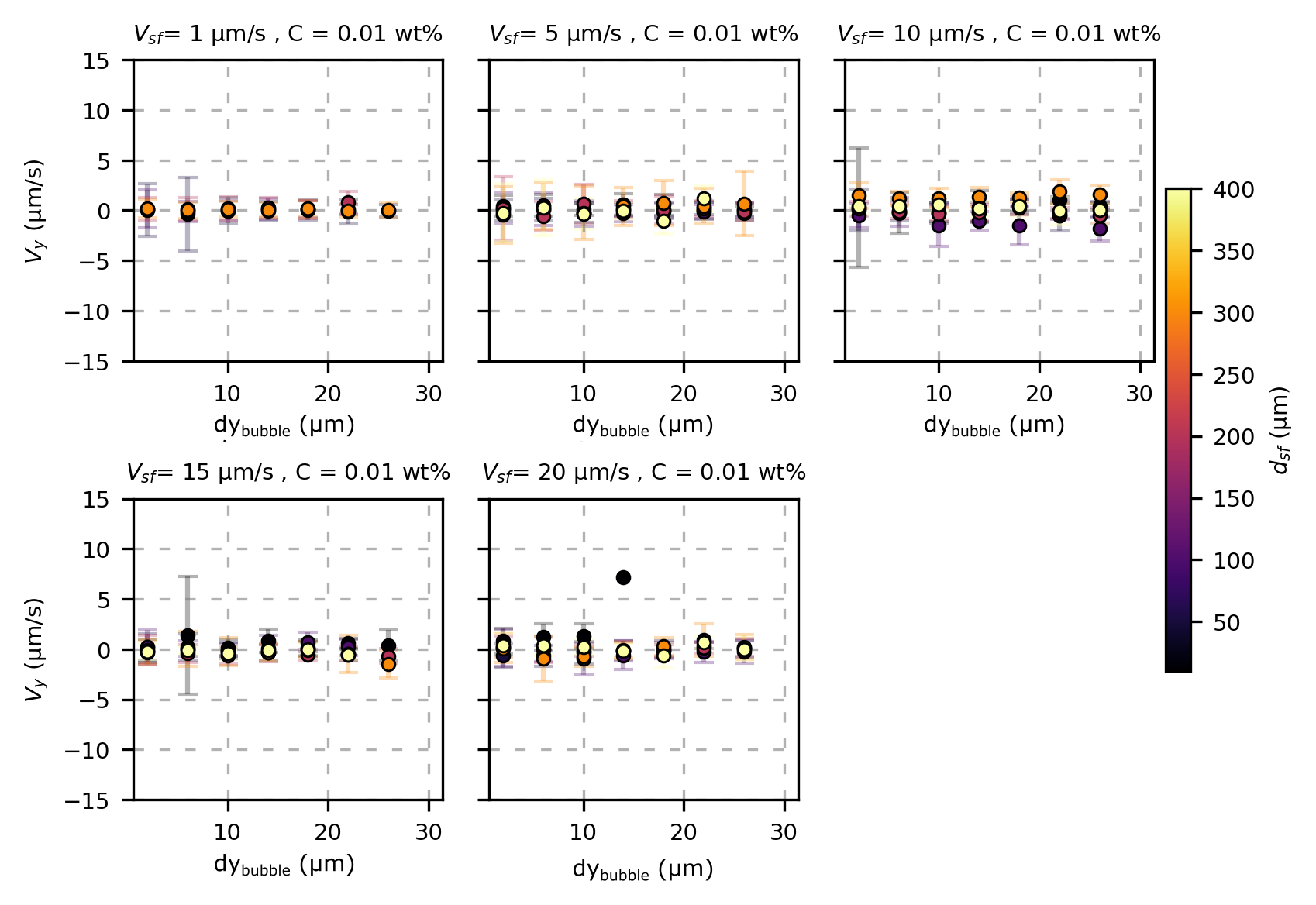}
 \caption{Mean fluid flow velocity $V_y$ near bubbles during liquid solidification in the direction orthogonal to the solidification front progression. The flow at the bubble's equator is assessed within a $\pm 3~\mu m$ range surrounding the $x_{\text{equator}}$ position and is expressed as a function of both the distance to the solidification front $d_{\text{sf}}$ and the distance from the bubble $y_{bubble}$. All experiments were performed using aqueous solutions containing $C = 0.01~wt.\%$ Tween 80 and with solidification velocities of 1, 5, 10, 15, and $20~\mu m/s$.}
 \label{fig:figure8}
\end{figure*}

Figure~\ref{fig:figure8} illustrates the velocity of fluid flow $V_y$ in a direction perpendicular to the solidification axis, displaying the separation to the equatorial bubble $y_{bubble}$ for varying distances $d_{\text{sf}}$ between the bubble and the solidification front. Across all examined solidification velocities, the average fluid flow velocities are consistently insignificant irrespective of the separation from the solidification front. However, the variability in measured vectors, as indicated by the standard deviation, increases when the proximity to the bubble reduces to under 10~$\mu m$.

\begin{figure*}
 \centering
 \includegraphics[width=14cm]{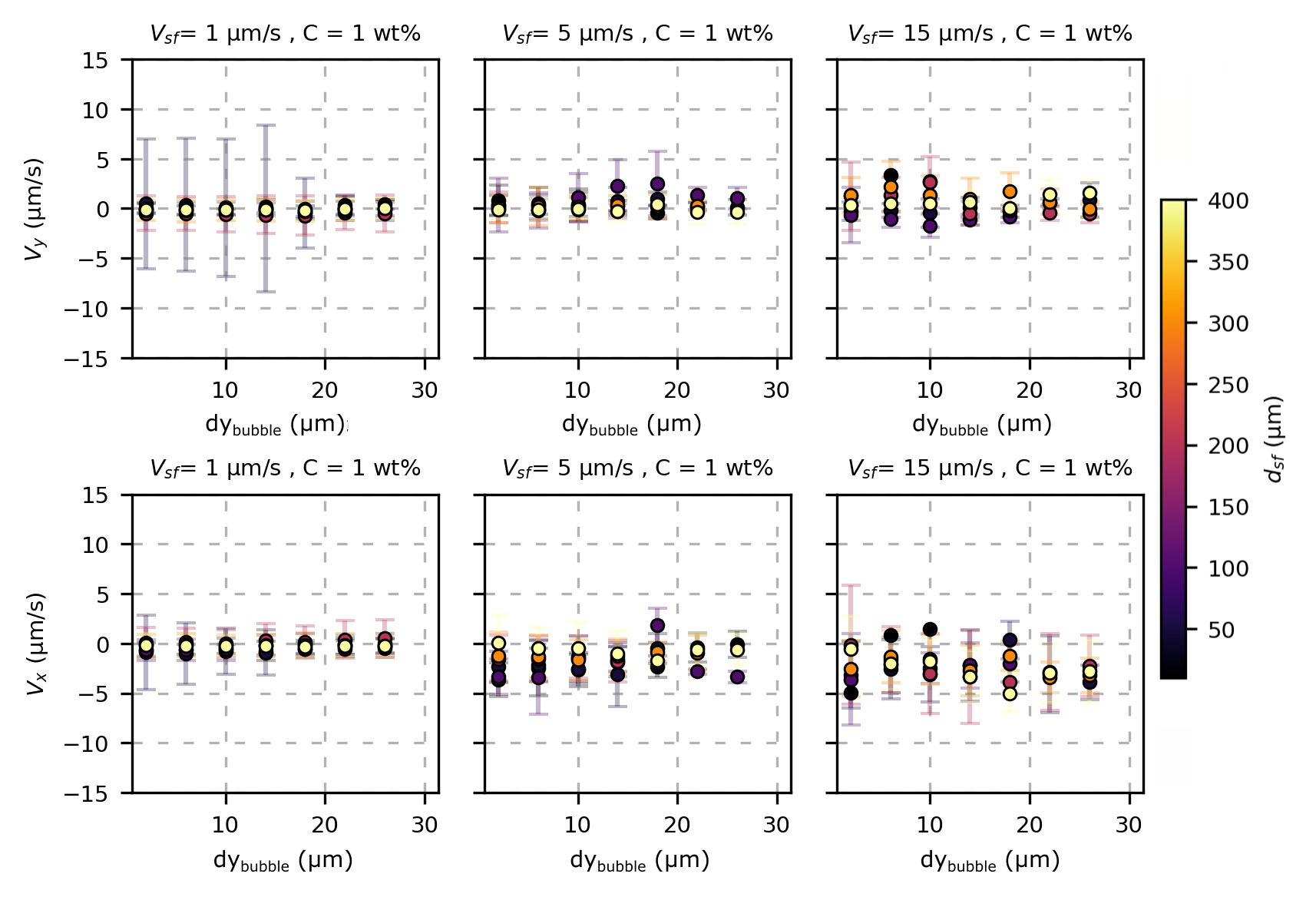}
 \caption{Mean fluid flow velocities $V_x$ and $V_y$ near bubbles during liquid solidification, in both directions, after subtraction of the flow induced by volumetric expansion, and for a surfactant concentration of 1~wt.\%. The flow is quantified at the equatorial region of the bubble within a $\pm 3~\mu m$ range centered around the $x_{\text{equator}}$ position. It is expressed as a function of both the distance from the solidification front and the distance from the bubble, measured in the direction orthogonal to the solidification front propagation ($V_y$, top row), as well as in the direction parallel to it ($V_x$, bottom row). All experiments were conducted using aqueous solutions comprising 1~wt.\% Tween 80, with a solidification velocity of 1, 5, and $15~\mu m/s$.}
 \label{fig:figure9}
\end{figure*}

We study the case of a higher surfactant concentration of $C = 1~wt.\%$ Tween 80. As illustrated in Fig.~\ref{fig:figure9}, we present the mean fluid flow velocities $V_x$ and $V_y$ in both directions during solidification, considering three distinct solidification velocities $V_{\text{sf}}$ ranging from 1 to 15~$\mu m$/s. It is observed that, across all three solidification velocities, the flow measurements in the solidification direction at 1~wt.\% (see to Fig.~\ref{fig:figure9}, bottom row) are comparable to those obtained at 0.01~wt.\%. These findings indicate that the surfactant concentration in the liquid does not significantly affect the flow dynamics around bubbles during the solidification process of our samples. Regrettably, due to the essential role of surfactants in bubble generation and stabilization, evaluating the effects of their absence on flow behavior was not feasible.

The findings presented thus far indicate that, within our system, Marangoni effects, although possibly present, do not substantially influence the liquid flow surrounding the bubbles, and the corresponding velocities generally do not exceed $5~\mu m/s$. 

We can estimate the order of magnitude of the velocities expected for Marangoni flows. Assuming that the motion of the bubbles in the interfacial tension gradient is controlled by viscous dissipation in the thin film between the top of the bubble and the top glass slide, we can use Bretherton’s model to estimate the velocities associated with Marangoni flows:

\begin{equation}
    U = \left(l_c\\\frac{d\gamma}{dx}\right)^{3/2} \frac{1}{\gamma^{1/2}\eta}
\end{equation}

where $l_c \approx 2\times10^{-3}~m$ is the capillary length, $\eta \approx10^{-3} Pa \cdot s$ the water viscosity, $\gamma\approx 50~mN/m$ the surface tension and $d\gamma/dx$ is the gradient of interfacial tension along the bubble surface. The variation in interfacial tension along the bubble surface can be inferred from the surface tension measurements detailed above. By considering an initial bubble diameter less than $100~\mu\text{m}$, both thermal and solutal gradients can be assessed at the scale of the bubble. For an applied temperature gradient of $15~^\circ\text{C/mm}$, the resultant temperature difference across a bubble with a diameter of $100~\mu\text{m}$ along the direction of solidification is $1.5~^\circ\text{C}$. As per the interfacial tension measurements depicted in Fig.~\ref{fig:figure2}, and assuming a linear relationship between surface tension and temperature, a bubble with a diameter of $100~\mu\text{m}$ would undergo a surface tension variation induced by thermal effects ranging from approximately 0.3 to 0.4~mN/m between its extremities. Taking
$d\gamma/dx = \frac{4 \times 10^{-4}}{10^{-4}} = 4~\text{N}\cdot\text{m}^{-2}$,
we should obtain velocities of the order of 1~m/s, which is several orders of magnitude higher than the velocities that we detect.

Furthermore, the interfacial tension gradient due to the concentration gradient is opposite to that of the temperature gradient. Indeed, the surfactant concentration is higher near the ice front; thus, the interfacial tension should decrease near the front. This effect may therefore induce motion in the opposite direction and potentially counteract the overall motion of the bubbles. Regarding surfactant gradients, the concentration profile near the ice front scales as follows~\cite{pohl1954solute}:

\begin{equation}
  \frac{C_{L}(x, t)-C_{0}}{C_{0}}=\left(\frac{1-k}{k}\right)\exp \left(-\frac{V_{s f}}{D} x\right)
  \label{eq:concentration_profile}
\end{equation}

with k the partition coefficient (the ratio of solute concentrations in the ice and in the bulk liquid), D the diffusion coefficient of solute in the liquid (in $m^2/s$), and $C_0$ the solute concentration (in g/l) far from the solidification front. The characteristic segregation length of the solute at the solidification front in a steady state regime can be evaluated as $L_d = D/V_{sf}$ ~\cite{tiller1953redistribution}. Under our experimental conditions, the characteristic length for micelles is estimated to be between $L_d = 1.5~\mu m$ for $V_{sf} = 20~\mu m/s$ and $L_d = 30~\mu m$ for $V_{sf} = 1~\mu m/s$~\cite{tyagi2022solute}. Assuming a partition coefficient $k\approx 0.1$, we should have a gradient of surfactant concentration of a factor 10 close to the front. Consequently, the bubble is anticipated to encounter a gradient in surface tension gradient whose magnitude would be comparable to thermal effects but oriented in the opposite direction. However, we note that at larger distances from the front, approximately on the order of $400~\mu m$, only thermal Marangoni flows are expected to be significant. Thus, the lack of a significant effect even far from the front suggests that both thermal and solutal Marangoni flows are negligible.

\section{Conclusions} 
\label{sec:conclusions}

This study demonstrates that volumetric expansion is the primary driver of fluid flow during the solidification of bubbly liquids, with velocities scaling linearly with the solidification rate, while Marangoni and phoretic effects play negligible roles under the tested conditions. Our findings challenge prevailing theoretical models that predict significant Marangoni flows during solidification~\cite{meijer2024bubble}, suggesting that these effects may be overestimated in micrometric systems or confined geometries. Instead, volumetric expansion emerges as the dominant mechanism, providing a simpler framework for predicting fluid dynamics in solidifying bubbly liquids. Future experimental explorations could reveal regimes where Marangoni effects are significant, bridging the gap between micrometric experiments and industrial-scale processes.

\section*{Conflict of interest} 
\label{sec:conflict_of_interest}
None.

\begin{acknowledgments}
We acknowledge the CNRS for its financial support of the COBLE project through the PRIME80 program, and the ANR - FRANCE (French National Research Agency) for its financial support of the FROST project ANR-22-CE06-0022.
\end{acknowledgments}

\textbf{CRediT author statement: }
\textbf{Bastien Isabella}: Investigation, Validation, Formal analysis, Writing - Original Draft, Visualization
\textbf{Emma Houllegatte}: Investigation, Validation
\textbf{Cécile Monteux}: Conceptualization, Writing - Review \& Editing, Supervision, Funding acquisition
\textbf{Sylvain Deville}: Conceptualization, Writing - Review \& Editing, Supervision, Funding acquisition

\bibliography{biblio.bib}

@misc{isabella2026,
      title={Coupled gas and bubble dynamics at the solidification front}, 
      author={Bastien Isabella and Cécile Monteux and Sylvain Deville},
      year={2026},
      eprint={2601.15045},
      archivePrefix={arXiv},
      primaryClass={cond-mat.soft},
      url={https://arxiv.org/abs/2601.15045}, 
}

@article{ahn2008co2,
  title={CO2 diffusion in polar ice: observations from naturally formed CO2 spikes in the Siple Dome (Antarctica) ice core},
  author={Ahn, Jinho and Headly, Melissa and Wahlen, Martin and Brook, Edward J and Mayewski, Paul A and Taylor, Kendrick C},
  journal={Journal of Glaciology},
  volume={54},
  number={187},
  pages={685--695},
  year={2008},
  publisher={Cambridge University Press}
}

@book{gow1977growth,
  title={Growth history of lake ice in relation to its stratigraphic, crystalline and mechanical structure},
  author={Gow, Anthony Jack and Langston, David},
  number={77},
  year={1977},
  publisher={Department of Defense, Army, Corps of Engineers, Cold Regions Research and~…}
}

@article{zhang2013nucleation,
  title={Nucleation, growth, transport, and entrapment of inclusions during steel casting},
  author={Zhang, Lifeng},
  journal={Jom},
  volume={65},
  number={9},
  pages={1138--1144},
  year={2013},
  publisher={Springer}
}

@article{li2013bubbles,
  title={Bubbles defects distribution in sapphire bulk crystals grown by Czochralski technique},
  author={Li, Hui and Ghezal, EA and Nehari, A and Alombert-Goget, Guillaume and Brenier, Alain and Lebbou, Kheirreddine},
  journal={Optical Materials},
  volume={35},
  number={5},
  pages={1071--1076},
  year={2013},
  publisher={Elsevier}
}

@article{bouaita2019seed,
  title={Seed orientation and pulling rate effects on bubbles and strain distribution on a sapphire crystal grown by the micro-pulling down method},
  author={Bouaita, Rekia and Alombert-Goget, G and Ghezal, EA and Nehari, A and Benamara, Omar and Benchiheub, M and Cagnoli, G and Yamamoto, K and Xu, X and Motto-Ros, Vincent and others},
  journal={CrystEngComm},
  volume={21},
  number={28},
  pages={4200--4211},
  year={2019},
  publisher={Royal Society of Chemistry}
}

@article{bronstein1981rejection,
  title={Rejection and capture of cells by ice crystals on freezing aqueous solutions},
  author={Bronstein, VL and Itkin, YA and Ishkov, GS},
  journal={Journal of Crystal Growth},
  volume={52},
  pages={345--349},
  year={1981},
  publisher={Elsevier}
}

@article{korber1988phenomena,
  title={Phenomena at the advancing ice--liquid interface: solutes, particles and biological cells},
  author={K{\"o}rber, Christoph},
  journal={Quarterly reviews of biophysics},
  volume={21},
  number={2},
  pages={229--298},
  year={1988},
  publisher={Cambridge University Press}
}

@article{ghezal2012observation,
  title={Observation of gas bubble incorporation during micropulling-down growth of sapphire},
  author={Ghezal, EA and Nehari, A and Lebbou, K and Duffar, T},
  journal={Crystal growth \& design},
  volume={12},
  number={11},
  pages={5715--5719},
  year={2012},
  publisher={ACS Publications}
}

@article{tyagi2022solute,
  title={Solute effects on the dynamics and deformation of emulsion droplets during freezing},
  author={Tyagi, Sidhanth and Monteux, C{\'e}cile and Deville, Sylvain},
  journal={Soft Matter},
  volume={18},
  number={21},
  pages={4178--4188},
  year={2022},
  publisher={Royal Society of Chemistry}
}

@article{tiller1953redistribution,
  title={The redistribution of solute atoms during the solidification of metals},
  author={Tiller, WA and Jackson, KA and Rutter, JW and Chalmers, Bruce},
  journal={Acta metallurgica},
  volume={1},
  number={4},
  pages={428--437},
  year={1953},
  publisher={Elsevier}
}

@article{pohl1954solute,
  title={Solute redistribution by recrystallization},
  author={Pohl, Robert G},
  journal={Journal of Applied Physics},
  volume={25},
  number={9},
  pages={1170--1178},
  year={1954}
}

@article{lubetkin1988nucleation,
  title={The nucleation of bubbles in supersaturated solutions},
  author={Lubetkin, Steven and Blackwell, Mark},
  journal={Journal of colloid and interface science},
  volume={126},
  number={2},
  pages={610--615},
  year={1988},
  publisher={Elsevier}
}

@article{geguzin1981crystallization,
  title={Crystallization of a gas-saturated melt},
  author={Geguzin, Ya E and Dzuba, AS},
  journal={Journal of Crystal Growth},
  volume={52},
  pages={337--344},
  year={1981},
  publisher={Elsevier}
}

@article{meulenbroek2021competing,
  title={Competing Marangoni effects form a stagnant cap on the interface of a hydrogen bubble attached to a microelectrode},
  author={Meulenbroek, AM and Vreman, AW and Deen, NG},
  journal={Electrochimica Acta},
  volume={385},
  pages={138298},
  year={2021},
  publisher={Elsevier}
}

@article{takagi2011surfactant,
  title={Surfactant effects on bubble motion and bubbly flows},
  author={Takagi, Shu and Matsumoto, Yoichiro},
  journal={Annual Review of Fluid Mechanics},
  volume={43},
  number={1},
  pages={615--636},
  year={2011},
  publisher={Annual Reviews}
}

@article{meijer2024bubble, 
title={Enhanced bubble growth near an advancing solidification front}, 
volume={996}, 
DOI={10.1017/jfm.2024.777}, 
journal={Journal of Fluid Mechanics}, 
author={Meijer, Jochem G. and Rocha, Duarte and Linnenbank, Annemarie M. and Diddens, Christian and Lohse, Detlef}, 
year={2024}, 
pages={A22}}

@article{young1959motion,
  title={The motion of bubbles in a vertical temperature gradient},
  author={Young, NO and Goldstein, Jo S and Block, MJ0087},
  journal={Journal of Fluid Mechanics},
  volume={6},
  number={3},
  pages={350--356},
  year={1959},
  publisher={Cambridge University Press}
}

@article{merritt1988migration,
  title={The migration of isolated gas bubbles in a vertical temperature gradient},
  author={Merritt, Randy M and Subramanian, R Shankar},
  journal={Journal of colloid and interface science},
  volume={125},
  number={1},
  pages={333--339},
  year={1988},
  publisher={Elsevier}
}

@article{zheng2002thermophoresis,
  title={Thermophoresis of spherical and non-spherical particles: a review of theories and experiments},
  author={Zheng, F},
  journal={Advances in colloid and interface science},
  volume={97},
  number={1-3},
  pages={255--278},
  year={2002},
  publisher={Elsevier}
}

@article{anderson1989colloid,
  title={Colloid transport by interfacial forces},
  author={Anderson, John L},
  journal={Annual review of fluid mechanics},
  volume={21},
  number={1},
  pages={61--99},
  year={1989},
  publisher={Annual Reviews 4139 El Camino Way, PO Box 10139, Palo Alto, CA 94303-0139, USA}
}

@article{lorenceau2006high,
  title={A high rate flow-focusing foam generator},
  author={Lorenceau, Elise and Sang, Yann Yip Cheung and H{\"o}hler, Reinhard and Cohen-Addad, Sylvie},
  journal={Physics of fluids},
  volume={18},
  number={9},
  year={2006},
  publisher={AIP Publishing}
}

@ARTICLE{Dedovets2018-tn,
  title     = "A temperature-controlled stage for laser scanning confocal
               microscopy and case studies in materials science",
  author    = "Dedovets, Dmytro and Monteux, Cécile and Deville, Sylvain",
  journal   = "Ultramicroscopy",
  publisher = "Elsevier B.V.",
  volume    =  195,
  number    = "August",
  pages     = "1--11",
  month     =  dec,
  year      =  2018,
  eprint    = "1712.08510",
  doi       = "10.1016/j.ultramic.2018.08.009",
  pmid      =  30172855,
  issn      = "0304-3991,1879-2723"
}

@ARTICLE{Schindelin2012-lt,
  title     = "Fiji: an open-source platform for biological-image analysis",
  author    = "Schindelin, Johannes and Arganda-Carreras, Ignacio and Frise,
               Erwin and Kaynig, Verena and Longair, Mark and Pietzsch, Tobias
               and Preibisch, Stephan and Rueden, Curtis and Saalfeld, Stephan
               and Schmid, Benjamin and Tinevez, Jean-Yves and White, Daniel
               James and Hartenstein, Volker and Eliceiri, Kevin and Tomancak,
               Pavel and Cardona, Albert",
  journal   = "Nat. Methods",
  publisher = "Nature Publishing Group, a division of Macmillan Publishers
               Limited. All Rights Reserved.",
  volume    =  9,
  number    =  7,
  pages     = "676--682",
  month     =  jun,
  year      =  2012,
  doi       = "10.1038/nmeth.2019",
  pmc       = "PMC3855844",
  pmid      =  22743772,
  issn      = "1548-7091,1548-7105"
}

@mic{openPIV,
  author       = {Alex Liberzon and
                  Theo Käufer and
                  Andreas Bauer and
                  Peter Vennemann and
                  Erich Zimmer},
  title        = {OpenPIV/openpiv-python: OpenPIV-Python v0.23.4},
  month        = jan,
  year         = 2021,
  publisher    = {Zenodo},
  version      = {0.23.4},
  doi          = {10.5281/zenodo.4409178},
  url          = {https://doi.org/10.5281/zenodo.4409178},
}

@article{mysels1990maximum,
  title={The maximum bubble pressure method of measuring surface tension, revisited},
  author={Mysels, Karol J},
  journal={Colloids and surfaces},
  volume={43},
  number={2},
  pages={241--262},
  year={1990},
  publisher={Elsevier}
}

@book{petrenko1999physics,
  title={Physics of ice},
  author={Petrenko, Victor F and Whitworth, Robert W},
  year={1999},
  publisher={OUP Oxford}
}

@article{piazza2008thermophoresis,
  title={Thermophoresis in colloidal suspensions},
  author={Piazza, Roberto and Parola, Alberto},
  journal={Journal of Physics: Condensed Matter},
  volume={20},
  number={15},
  pages={153102},
  year={2008},
  publisher={IOP Publishing}
}

@book{degroot2013,
  title={Non-equilibrium thermodynamics},
  author={De Groot, Sybren Ruurds and Mazur, Peter},
  year={2013},
  publisher={Courier Corporation}
}

@article{duhr2006molecules,
  title={Why molecules move along a temperature gradient},
  author={Duhr, Stefan and Braun, Dieter},
  journal={Proceedings of the National Academy of Sciences},
  volume={103},
  number={52},
  pages={19678--19682},
  year={2006},
  publisher={National Academy of Sciences}
}

@article{braibanti2008does,
  title={Does thermophoretic mobility depend on particle size?},
  author={Braibanti, Marco and Vigolo, Daniele and Piazza, Roberto},
  journal={Physical review letters},
  volume={100},
  number={10},
  pages={108303},
  year={2008},
  publisher={APS}
}

  \section{Appendixes}
Table I summarizes the experimental conditions and sample compositions used to determine their influence on flow patterns in the liquid phase. The surfactant used is Tween 80, with concentration C ranging from 0.001 to 1~wt.\%. Experiments were performed across a range of temperature gradients (0--25$~^\circ \text{C/mm}$) and solidification velocities $V_{\text{sf}}$= 0--20$~\mu m/s$. The final entry in the table depicts a standard experiment involving the solidification of a bubbly liquid containing surfactant. 

\begin{turnpage}
\begin{table}
\begingroup
\squeezetable
\begin{ruledtabular}
\begin{tabular}{c|c|c|c|c|c|c|c|c|c|c|c|c}
 \textbf{Conditions} &\multicolumn{4}{c|}{\cellcolor{lightgray}\textbf{Parameters}}&\multicolumn{8}{c}{\textbf{Transport mechanisms}}\\\hline
 & Surfactant & $T^\circ$  & Freezing  & Bubbles & Brownian  & Tracers  & Micelles  & Volumetric  & Tracers  & Bubbles  & Thermal  & Solutal  \\
 & & gradient & front & & motion & thermophoresis & thermophoresis & expansion & diffusiophoresis & diffusiophoresis & Marangoni & Marangoni \\ \hline
\cellcolor{lightgray} 1 &   &   &  &   & x &   &  &  &  &  &  &      \\ \hline
\cellcolor{lightgray} 2 & x &   &  &   & x &   &  &  &  &  &  &      \\ \hline
\cellcolor{lightgray} 3 & x &   &  & x & x &   &  &  &  &  &  &      \\ \hline
\cellcolor{lightgray} 4 &   & x &  &   & x & x &  &  &  &  &  &      \\ \hline
\cellcolor{lightgray} 5 & x & x &  &   & x & x & x &  &  &  &  &      \\ \hline
\cellcolor{lightgray} 6 & x & x &  & x & x & x & x &  &  &  & x &      \\ \hline
\cellcolor{lightgray} 7 &  & x & x &  & x & x &  & x & x &  &  &      \\ \hline
\cellcolor{lightgray} 8 & x & x & x &  & x & x & x & x & x &  &  &      \\ \hline
\cellcolor{lightgray} 9 & x & x & x & x &x  & x & x & x & x & x &x  & x     \\
\end{tabular}
\caption{}
\end{ruledtabular}
\endgroup
\label{tab:conditions}
\end{table}
\end{turnpage}

\end{document}